\documentclass[a4paper,11pt]{article}
\pdfoutput=1 

\usepackage{jheppub} 
\usepackage{empheq}
\usepackage{bm}
\usepackage{slashed}
\usepackage[T1]{fontenc} 
\usepackage[all]{xy}
\usepackage{rotating}

\def\cN{\mathcal{N}}

\usepackage{slashed}

\def\ep{\epsilon}

\def\CN{{\cal N}}
\def\CO{{\cal O}}

\def\BZ{{\mathbb Z}}

\def\s{\sigma}

\def\){\right)}
\def\({\left( }
\def\]{\right] }
\def\[{\left[ }

\def\beq#1\eeq{\begin{align}#1\end{align}}

\newcommand{\bea}{\begin{eqnarray}}
\newcommand{\eea}{\end{eqnarray}}
\newcommand{\ra}{\rightarrow}
\newcommand{\zt}{{\tilde z}}

\newcommand{\nn}{\nonumber}

\newcommand{\z}{\rho}

\title{\boldmath Solving Mass-deformed Holography Perturbatively}

\author{Nakwoo Kim}

\affiliation{Department of Physics and Research Institute of Basic Science,
	Kyung Hee University, Seoul 02447, Republic of Korea}
\emailAdd{nkim@khu.ac.kr}

\abstract{We study supergravity BPS equations which correspond to mass-deformation of some representative AdS/CFT examples. The field theory of interest are ${\cal N}=4, D=4$ super Yang-Mills, the ABJM model in $D=3$, and the Brandhuber-Oz fixed point in $D=5$. For these gauge theories the free energy with mass terms for matter multiplets is calculable in large-$N$ limit using supersymmetric localization technique. We suggest a perturbative method to solve the supergravity equations. For the dual of mass-deformed ABJM model we reproduce the known exact solutions. For the mass-deformed Brandhuber-Oz theory our method gives the holographic free energy in analytic form. For ${\cal N}=2^*$ theory our result is in good agreement with the localization result.}

\begin{document} 
\maketitle
\flushbottom

\section{Introduction}
\label{sec:intro}
According to the AdS/CFT correspondence \cite{Maldacena:1997re}, classical solutions in supergravity theories which have string theory origin carry exact information on strongly coupled dual field theory in large-$N$ limit. Because the construction of analytic solutions in supergravity is relatively easier than the evaluation of path integral in strongly-coupled regime, in early days of AdS/CFT we were given lots of results from the supergravity side, most of which just had to wait until the strongly-coupled field theory result is available. Such {\it predictions} of supergravity include central charge, partition function, and most generally correlation functions. 

We now see the tide is turned, with 
the recent proliferation of localization results \cite{Pestun:2007rz} for supersymmetric field theories in various dimensions. In simple terms, the Newton constant of lower-dimensional AdS (gauged) supergravity gives us the central charge/partition function of the dual field theory. Confirmation of a long-time mystery in M-theory, namely the $N^{3/2}$ scaling of degrees of freedom, with the correct numerical coefficients, has been achieved for various three-dimensional super-conformal field theories which are realized on M2-branes \cite{Kapustin:2009kz,Jafferis:2010un,Hama:2010av,Drukker:2010nc,Herzog:2010hf,Martelli:2011qj,Cheon:2011vi,Jafferis:2011zi}. We also note that the supergravity solution for M5-branes wrapped on 3-cycles and $N^3$ scaling of the free energy was confirmed with correct coefficients against field theory computation \cite{Gang:2014qla,Gang:2014ema}, thanks to the localization results and 3d/3d correspondence reported in \cite{Lee:2013ida,Cordova:2013cea}. In addition to $D=4$ and $D=3$ theories, in this paper we will also deal with the duality of $D=5$ gauge theories \cite{Brandhuber:1999np,Jafferis:2012iv}, in particular the ones realized in massive IIA theory with $N^{5/2}$ scaling of the degrees of freedom. 

Although the free energy in the super-conformal phase has been successfully matched for a number of AdS/CFT examples, the localization computation actually provides more information than that. The point is, the localization prescription goes through even with non-zero mass of matter fields. On the gravity side it leads to a domain wall solution with non-trivial scalar fields. In other words, while the extension to non-conformal phase on the field theory side is rather straightforward (at least for certain mass-deformations), the dual description involves solving quite a non-trivial set of BPS equations in supergravity. 

This program, which we may call mass-deformed precision holography, has been tackled for several years by now. Freedman and Pufu \cite{Freedman:2013ryh} considered the subsector of maximal $SO(8)$ gauged supergravity in four-dimensional Euclidean space with scalars dual to mass-terms of the ABJM model \cite{Aharony:2008ug}. They found exact solutions to the BPS equations, and have shown that the holographically renormalized action is consistent with the localization calculation of partition function with non-zero mass, in large-$N$ limit. Then 
Bobev et al. \cite{Bobev:2013cja} correctly identified the analogous sub-sector of ${\cal N}=8, \, SO(6)$ gauged supergravity in five-dimensional Euclidean signature, presented the BPS equations, and argued that their numerical solutions are consistent with the prediction of localization result. Recently this program was also applied to a duality pair between $D=5$ gauge theory and six-dimensional supergravity, in \cite{Chang:2017mxc,Gutperle:2018axv}. More work on the mass deformation of other AdS/CFT duality pairs can be found in  \cite{Karndumri:2012vh,Balasubramanian:2013esa,Bigazzi:2013xia,Karndumri:2014lba,Karch:2015kfa,Chen-Lin:2015xlh,Bobev:2016nua,Kol:2016ucd,Gutperle:2017nwo,Kim:2018sdw,Bobev:2018hbq,Bobev:2018eer,Bobev:2018wbt,Russo:2019lgq}.

We emphasize that except for the mass-deformed BPS solutions of the ABJM model, for gravity dual \cite{Bobev:2018hbq} of ${\cal N}=2^*$ mass-deformation of $D=4, \, {\cal N}=4$ super Yang-Mills theory and the dual \cite{Gutperle:2018axv} of mass-deformed Brandhuber-Oz duality, only numerical solutions were constructed. That clearly implies there is room for improvement. In this paper we aim to fill this gap and concentrate on solving the BPS equations. Our idea is to introduce a perturbative parameter and linearize the BPS equations. The BPS equations are, involving scalar fields with a complicated potential, nonlinear. We will see that, using the perturbation trick, at leading non-trivial order we have a set of nonlinear but {\it homogeneous} equations, and if we manage to solve them, at higher orders we have linear differential equations. Then solving the BPS equation just reduces to performing integrals. In Sec.\ref{ads4} we re-visit the BPS equations for mass-deformation of the ABJM model, and solve them using our method to show its power. In Sec.\ref{ads6} we solve perturbatively the BPS equations for mass-deformed Brandhuber-Oz theory, as presented in  \cite{Gutperle:2018axv}. In particular, we re-sum the series form of a central relation between integration constants of the BPS equations and present an analytic form of the holographic free energy in terms of the mass parameter. In Sec.\ref{ads5}, we re-visit the BPS equations of ${\cal N}=2^*$ holography, presented in \cite{Bobev:2018hbq}. In this case, due to $\log$-behavior near UV, explicit integration becomes rather messy. In this paper we use approximation in terms of a series expansion near IR, to solve our perturbative equations. We will show that our result is in good agreement with the conjecture made in \cite{Bobev:2018hbq}, and thus also with the localization result. We conclude in Sec.\ref{discussion}.

We note that perturbative approaches have been already employed successfully to other problems in AdS/CFT in different context such as black holes and deformation of boundary metric, in \cite{Bhattacharyya:2010yg,Alday:2014rxa,Alday:2014bta}.

\section{$AdS_4$: mass deformations in ABJM}
\label{ads4}
Let us first consider the ABJM theory \cite{Aharony:2008ug} as a warm-up. On the gauge theory side we have $\cN=6, \,\, D=3$ Chern-Simons matter theory, and on the gravity side we have $AdS_4\times S^7/\BZ_k$. 
The Chern-Simons theory has a quiver structure and $U(N)\times U(N)$ gauge groups have level $(k,-k)$ which gives $\BZ_k$ orbifolding of the vacuum moduli space. 

The relevant Einstein-scalar system of the holographic side in Euclidean signature is derived, as a truncation from $\cN=8$, $D=4$ gauged supergravity, in \cite{Freedman:2013ryh}. Eq.(4.1) of this reference reads
\begin{align}
\label{action4}
S &= \frac{1}{8\pi G_4} \int d^4 x \sqrt{g} \left[ -\frac{1}{2}R + \sum_{\alpha=1}^3 \frac{\partial_\mu z^\alpha \partial^\mu \tilde z^\alpha}{(1-z^\alpha \tilde z^\alpha)^2} + \frac{1}{L^2} 
\left( 3 - \sum_{\alpha=1}^3 \frac{2}{1-z^\alpha \tilde z^\alpha}\right)
\right]
.
\end{align}
We use the following form of the metric\footnote{Compared to the convention of \cite{Freedman:2013ryh}, we use the conformal gauge $e^{B}=e^{A}/r$ and set $L=1$.},
\beq
\label{metric4}
ds^2 &= e^{2A(r)} (dr^2/r^2 +ds^2_{S^3} ) . 
\eeq
Note that the pure anti-de Sitter case (or the hyperbolic space here since we consider the Euclidean version) with radius $L$ corresponds to 
\beq
\label{pure}
e^{2A} = \frac{4r^2L^2}{(1-r^2)^2} . 
\eeq
Note that with this parametrization the range of holographic coordinate $r$ is $0\le r \le1$.
Upon change of variable $r=\tanh (u/2)$, we have an alternative form of AdS metric,
\beq
ds^2 = L^2 (du^2 + \sinh^2 u \, ds^2_{S^3} )\, . 
\eeq

It is also assumed that all scalar fields $z^\alpha, \tilde z^\alpha$ (which are dual to bosonic/fermionic mass terms) are functions of $r$ only and we keep the isometry of the round $S^3$. Then the BPS equations which solve the second-order field equations from \eqref{action4} with ansatz \eqref{metric4} are (See eq. (5.20) of \cite{Freedman:2013ryh}, and we set $L=1$.)
\beq
\label{bps4}
r(1+  \tilde z^1 \tilde z^2 \tilde z^3 ) z^{\alpha\prime} &=
(\pm 1 -r A' )(1-z^\alpha \tilde z^\alpha )
\left( z^\alpha + \frac{\tilde z^1 \tilde z^2 \tilde z^3}{\tilde z^\alpha} \right) , 
\nn
\\
r(1+   z^1  z^2  z^3 ) \tilde z^{\alpha\prime} &=
(\mp 1 -r A' )(1-z^\alpha \tilde z^\alpha )
\left( \tilde z^\alpha + \frac{z^1 z^2 z^3}{z^\alpha} \right) ,
\\
-1 &= -r^2 (A^\prime)^2 +e^{2A}\frac{(1+z^1 z^2 z^3)(1+\tilde z^1 \tilde z^2 \tilde z^3)}{\prod_{\beta=1}^{3} 
(1-z^\beta \tilde z^\beta)} . 
\nn
\eeq

In fact, although the above equations look rather involved and non-linear, exact solutions can be given in a rather simple analytic form. One can choose, for concreteness, the upper signs above and verify
\beq
z^\alpha(r) = c_\alpha f(r), \quad 
\tilde z^\alpha (r) = -\frac{c_1c_2c_3}{c_\alpha} f(r)
.
\eeq
Here $c_\alpha$ are three integration constants and all six scalar fields turn out to share the same profile with
\beq
\label{solsc}
f(r) = \frac{1-r^2}{1+c_1c_2c_3 r^2} . 
\eeq
And finally the conformal factor of the metric is 
\beq
\label{solme}
e^{2A} = \frac{4r^2(1+c_1c_2c_3)(1+c_1c_2c_3 r^4)}{(1-r^2)^2(1+c_1c_2c_3r^2)^2} . 
\eeq

Using the above configuration, the authors of \cite{Freedman:2013ryh} carefully identified how $c_\alpha$ are related to  mass parameters in the dual gauge theory and confirmed that the holographically renormalized action correctly leads to the $S^3$ partition function of the ABJM theory with mass terms. In this work, we are not going to add anything to the part of holographic computation. Instead of that, using $c_\alpha$ as small parameters, we will perform a perturbative analysis to study the behavior of regular gravity solutions. Let us emphasize here that our perturbative expansion is different from the usual near-UV expansion, and at each order of the mass parameter $c_\alpha$ we will solve the differential equations exactly. What we expect to obtain is clear: assuming that $c_\alpha$ are all small and ${\cal O}(\ep)$ for an auxiliary parameter $\ep$, we should obtain
\beq
f(r) &= (1-r^2) - c_1c_2c_3 r^2(1-r^2) + c_1^2 c_2^2 c_3^2 r^4 (1-r^2) +\cdots , 
\\
e^{2A} &= \frac{4r^2}{(1-r^2)^2} \left( 1 +  c_1c_2 c_3  (1-r^2)^2 + 2 r^2 c_1^2 c_2^2 c_3^2 (1-r^2)^2+ \cdots \right). 
\label{e2aexp}
\eeq

Now let us pretend we do not know the exact solutions. We write the unknown functions as follows and substitute them into the BPS equations.
\beq
z^\alpha (r) &= \sum_{k=1}^\infty \ep^k z^\alpha_k (r) \, , \quad
\tilde z^\alpha (r)= \sum_{k=1}^\infty \ep^k \tilde z^\alpha_k (r) \, , 
\\
e^{2A(r)} &= \frac{4r^2}{(1-r^2)^2} \left( 1 + \sum_{k=1}^\infty \ep^k a_k (r) 
\right) \, . 
\eeq

At zeroth order of $\ep$, we have the pure anti-de Sitter vacuum solution and the first two equations of \eqref{bps4} are trivial. The third equation is satisfied with eq. \eqref{pure} and vanishing scalar fields. Now at ${O}(\ep)$, the first two BPS equations give (for each $\alpha=1,2,3$)
\beq
\label{p1}
z^{\alpha\prime}_1 +\frac{2r}{1-r^2} z_1^\alpha &= 0 , 
\\
\label{p1t}
\tilde z^{\alpha\prime}_1 +\frac{2}{r(1-r^2)} \tilde z_1^\alpha &= 0 .
\eeq
One can easily see that $z^\alpha_1$ are proportional to $(1-r^2)$, while $\tilde z^\alpha_1$ are proportional to $1-r^{-2}$. Since we want a regular solution in the range of $0\le r \le 1$, we should set $\tilde z^\alpha_1 = 0$ and conclude at first order of the perturbation $z^\alpha_1=c_\alpha (1-r^2)$. Now we treat $c_\alpha\ep$ as an ${O}(\ep)$ quantity, and then plug $z^\alpha_1$ back to the second BPS equation and find 
\beq
\label{p2}
\tilde z^{\alpha\prime}_2 + \frac{2}{r(1-r^2)} \tilde z_2^\alpha = -\frac{2 c_1c_2 c_3(1-r^2)}{c_\alpha r}  \, . 
\eeq
This is again easily solved, and imposing regularity one gets
\beq
\tilde z^\alpha_2 = - \frac{c_1 c_2 c_3}{c_\alpha} (1-r^2) . 
\eeq
We now consider equations at ${O}(\ep^3)$. There appear equations for $z^\alpha_3,\tilde z^\alpha_3$, which satisfy exactly the same equations as \eqref{p1} and \eqref{p1t}. $\tilde z^\alpha_3$ should be again set to zero due to regularity. On the other hand, $z^\alpha_3 \sim (1-r^2)$ is allowed, but in fact this can be removed by re-definition of $c_\alpha$ and we can set $z^\alpha_3=0$. The last equation in \eqref{bps4} now gives an in-homogeneous equation for $a_3$,
\beq
(1-r^4)a_3' - 4r a_3 +4 c_1 c_2 c_3 r(2+r^2)(1-r^2)^2 = 0 . 
\eeq
When we solve this equation with the regularity condition at $r=1$, we obtain
\beq
a_3 = c_1 c_2 c_3 (1-r^2)^2 ,
\eeq
which agrees with the expansion of \eqref{e2aexp}.

This procedure can be obviously repeated to higher orders of $\ep$: the general feature is that at the leading order of the perturbation parameter, we have homogeneous first order differential equations \eqref{p1}, and we introduce integration constants which will be treated as small perturbation parameters. Then at higher orders, the equations are in-homogenious like \eqref{p2} but due to regularity condition the solution is unique at each order, since homogeneous solutions are generically singular. This is true in equations for $\tilde z^\alpha, A$ and in fact at $\ep^4,\ep^7$ etc there is freedom to add homogeneous solutions of the leading order. But this freedom can be eliminated by re-definition of $\ep$, and un-physical. One can indeed check that the perturbative solutions agree with the series-expanded terms of \eqref{solsc} and \eqref{solme}.

Before we turn to the next example, let us add a comment on the advantage of using the metric \eqref{metric4}. Usually in the numerical study of holographic flow, the metric ansatz of $ds^2= du^2 + e^{2A(u)} ds^2_{S^3}$ would be preferred. In that parametrization, for general solutions the location of IR $u=u_{IR}$ (where $e^{2A}$ vanishes) is not at $u=0$ any more, and $u_{IR}$ varies as a function of UV parameters. However, because our BPS equations \eqref{bps4} have no shift symmetry in $r$, and as one can easily verify from explicit integration of our perturbative equations, the range of $r$ remains in $0\le r\le 1$ even for $\ep\neq 1$. At every order of $\ep$, one demands regularity at $r=0$ (IR) and $r=1$ (UV).
\section{$AdS_6$: Mass deformation of $D=5$ SCFT}
\label{ads6}
Let us now turn to the example of $AdS_6/CFT_5$. The example of our interest is the solution in $F(4)$ gauged supergravity \cite{Romans:1985tw}. Unlike other examples we consider in this paper, this theory is not maximally supersymmetric and the dual field theory also has just half of maximal supersymmetry. For a long time the higher dimensional origin of this supergravity was known only for massive IIA string theory, and the AdS vacuum is due to a configuration of D4-branes, D8-branes, and O8-planes \cite{Brandhuber:1999np,Cvetic:1999un}. The quantitative comparison between the supergravity and gauge theory has been made first in \cite{Jafferis:2012iv}, using the localization formula. We note that recently other uplifting prescriptions to IIB supergravity have been constructed in \cite{Hong:2018amk,Malek:2018zcz,Malek:2019ucd}, along the direction of \cite{Apruzzi:2014qva,Kim:2015hya,DHoker:2016ujz,DHoker:2017mds,DHoker:2017zwj}. 

The associated mass deformations are studied more recently in \cite{Chang:2017mxc,Gutperle:2018axv}. In this section we revisit the computation of Ref. \cite{Gutperle:2018axv}, where the authors considered adding vector multiplets to $F(4)$ supergravity and turned on scalar fields thereof, in order to consider mass deformation of matter fields on the gauge theory side. 

Let us repeat the formulae for the action and the BPS equations as given in \cite{Gutperle:2018axv}. There will be a slight difference in the presentation of BPS equations, since we choose a different convention for the metric and the holographic coordinate in order to make our perturbative results look simpler. 

The relevant gravity action in Euclidean signature when truncated to Einstein-scalar sector is 
\beq
S = \frac{1}{4\pi G_6} \int d^6x \sqrt{g} 
\left(
-\frac{1}{4} R + \partial_\mu \sigma \partial^\mu \sigma 
+ \frac{1}{4} G_{ij} (\phi) \partial_\mu \phi^i \partial^\mu \phi^j
+ V(\sigma, \phi^i)
\right)
\eeq
With   $\sigma$ and $\phi^i \, (i=0,1,2,3)$, we have in total five scalar fields here. The metric on the scalar manifold parametrized by $\phi^i$ is 
\beq
G_{ij} = \mbox{diag} ( \cosh^2 \phi^1 \cosh^2 \phi^2 \cosh^2 \phi^3 , 
\cosh^2 \phi^2 \cosh^2 \phi^3 , \cosh^2 \phi^3 , 1
) . 
\eeq
And the scalar potential is rather complicated, 
\begin{align}\label{scalpota}
V(\sigma,\phi^i) =&- g^2 e^{2 \sigma }+\frac{1}{8} m e^{-6 \sigma } \bigg[-32 g e^{4 \sigma } \cosh \phi^0 \cosh \phi^1 \cosh \phi^2 \cosh \phi^3+8 m \cosh ^2\phi^0
\nn\\
&+m \sinh ^2 \phi^0 \bigg(-6+8 \cosh ^2\phi^1 \cosh ^2 \phi^2 \cosh (2 \phi^3)+\cosh (2 (\phi^1-\phi^2))\nn\\
&+\cosh (2 (\phi^1+\phi^2))+2 \cosh (2 \phi^1)+2 \cosh (2 \phi^2)\bigg)\bigg]
\end{align}
For the supersymmetric case $g=3m$ and the AdS vacuum has unit radius for $m=1/2$.

We choose the metric convention just as in the previous section, and
\beq
\label{conv6}
ds^2 =  e^{2A(r)} \left(dr^2/r^2 + ds^2_{S^5}\right) . 
\eeq
All scalar fields are again functions of $r$ only. 
From the supersymmetry transformation rules, the BPS equations can be derived.  With the metric convention of eq. \eqref{conv6}, they are as follows.
\beq
r e^{-A} A' &= 2 (G_0 S_0 +G_3 S_3 ) 
\label{bps6a}
\\
r e^{-A}\sigma' &= 2\eta \sqrt{ N^2_0 + N^2_3}
\label{bps6b}
\\
r e^{-A}\cos \phi^3 (\phi^0)' &= - (G_0 M_0 + G_3 M_3 )
\label{bps6c}
\\
r e^{-A}(\phi^3)' & = i ( G_3 M_0 - G_0 M_3 )
\label{bps6d}
\eeq
Additionally there is an algebraic relation which is also a consequence of supersymmetry. 
\beq
e^{-2A} = 4 (G_0 S_0 + G_3 S_3 )^2 - 4 ( S^2_0 + S^2_3) . 
\label{algcon}
\eeq
Note that here we set $\phi^1=\phi^2=0$, which is a consistent truncation of the action and in fact a requirement of an unbroken R-symmetry \cite{Gutperle:2018axv}. The expressions on the right-hand-side  above include several new symbols which are introduced in order to make the equations concise. 
\bea
\label{newdefs}
G_0 &=& \eta \frac{N_0}{\sqrt{N^2_0+N^2_3}} \nn\\
G_3 &=& -\eta \frac{N_3}{\sqrt{N^2_0+N^2_3}} \nn\\
S_0&=&\frac14 \left(g\cos \phi^3 e^\s+m e^{-3\s}\cosh \phi^0\right)
\nn\\
S_3&=&\frac14 i \,m ~e^{-3\s}\sinh \phi^0 \sin \phi^3
\\
N_0&=&-\frac14 \left(g\cos \phi^3 e^\s-3m e^{-3\s}\cosh \phi^0\right)
\nn\\
N_3&=&-\frac34 i\, m e^{-3\s}\sinh \phi^0 \sin \phi^3
\nn\\
M_0&=&2m ~e^{-3\s}\cos \phi^3\sinh \phi^0
\nn\\
M_3&=&-2 i\, g ~e^{\s}\sin \phi^3 \nn
\eea
Here $\eta=\pm1 $ and the choice is related to the range of $r$: for $0\le r \le 1$ we need to choose $\eta=-1$.

In order to obtain the partition function from gravity solutions, one needs to understand the UV expansion. In \cite{Gutperle:2018axv}, the authors used the standard convention of Fefferman-Graham coordinates for asymptotic AdS and chose the metric $ds^2 = d\z^2/\z^2 + e^{2f(\z)} ds^2_{s^5}$. The result of the series expansion for small $\z$ is then  
\beq
f & = - \log \z + f_k - \left(\frac{1}{4} e^{-2f_k} + \frac{1}{16} \alpha^2 \right)
\z^2 + O(\z^4) \, , 
\label{uv6a}
\\
\sigma &= \frac{3}{8} \alpha^2 \z^2 + \frac{1}{4} e^{f_k} \alpha \beta \z^3 + O(\z^4) \, , 
\label{uv6b}
\\
\phi^0 &= \alpha \z - \left( \frac{5}{4} \alpha e^{-2f_k} + \frac{23}{48} \alpha^3 \right) \z^3
+ O(\z^4) \, , 
\\
\phi^3 &= e^{-f_k} \alpha \z^2 + \beta \z^3 + O(\z^4) \, . 
\label{uv6d}
\eeq
Note that for pure AdS case we have a relation $\z=(1-r)/(1+r)$ and $A=f$ between the two different coordinate choices. When we turn off the scalar fields completely, $\alpha=\beta=0$, $e^A= \frac{1-\z^2}{2\z}$, and $f_k=-\log 2$. Obviously, in perturbative approach $\alpha, \beta,$ and $2e^{f_k}-1$ should be treated as small parameters.

We have less scalar fields (three) here than the previous example of mass-deformed ABJM model (six), but the BPS equations are apparently more complicated. In fact, the authors of \cite{Gutperle:2018axv} resorted to numerical construction of regular solutions. It turns out that as one imposes regularity in IR ($r\ra 0$) there exists a one-parameter family of solutions. That means for instance $\beta,f_k$ are determined as functions of $\alpha$. According to the derivation in \cite{Gutperle:2018axv}, the holographically calculated free energy (the finite part of the on-shell supergravity action with Gibbons-Hawking and counter-terms included) is 
\beq
\frac{dF}{d\alpha} = \frac{\pi^2}{8 G_6}\beta e^{4f_k}\left( 4 - \alpha \frac{df_k}{d\alpha} \right) . 
\label{fprime}
\eeq
Then one can calculate the gravitational-side free energy $F(\alpha)$ by integration and combining it with the previous result for free energy of the superconformal field theory without mass deformation, {\it i.e.} $F(\alpha=0)$ \cite{Jafferis:2012iv} in large $N$ limit.
The free energy without mass terms is 
\beq 
F(\alpha=0) &= -\frac{9\sqrt{2}\pi N^{5/2}}{5\sqrt{8-N_f}} .
\eeq
Note that in this process the identification of Newton constant in terms of the brane number $N$ is essential:
\beq 
G_6 = \frac{5\pi \sqrt{8-N_f} }{27\sqrt{2}} N^{-5/2} . 
\eeq

So far we have reviewed the setup of \cite{Gutperle:2018axv}, and now let us turn to our perturbative analysis. The UV asymptotic behavior above implies that at the leading order of perturbation we should turn on $\phi^0,\phi^3$, while $\sigma$ begins to appear at second order. Just as we did in the previous section, we introduce a parameter $\ep$ and write
\beq
\begin{aligned}
\phi^0 (r)&= \sum_{k=1}^\infty \ep^k p_k (r)
\, , \quad
\phi^3 (r) = \sum_{k=1}^\infty \ep^k q_k (r) , 
\\ 
\sigma (r) &= \sum_{k=2}^\infty \ep^k \s_k (r) , 
\\
e^{A(r)} &= \frac{2r}{1-r^2} \left( 1 + \sum_{k=2}^\infty \ep^k a_k (r) \right).
\end{aligned}
\eeq

Substituting them into the BPS equations, we get infinitely many differential equations, one for each of $p_k, q_k,\s_k, a_k$. At the leading nontrivial order, we obtain a system of homogeneous and non-linear first order differential equations for $p_1,q_1,\s_2$. We find it convenient to consider $s_2$ so that 
\beq
\s_2 
= (2 s_2+p_1^2+q_1^2)/8 \, . 
\label{fors2}
\eeq
Then from the first non-trivial order expansion of the BPS equations above, we have the following conditions.
First, from the ${O}(\ep^0)$ part of the first BPS equation \eqref{bps6a}, we have (one also has to assume $p_1,q_1,s_2$ are all positive, without losing generality)
\beq
s_2 = \left(\frac{1+r^2}{1-r^2}\right) p_1 q_1 \, .
\label{relspq}
\eeq
We then consider the ${O}(\ep)$ equations of \eqref{bps6c}, \eqref{bps6d} and making use of \eqref{relspq} we get
\beq
p_1'(r) &= \frac{-(1+r^2)p_1(r)+3(1-r^2)q_1(r)}{r(1-r^2)} \, , 
\label{6heq1}
\\
q_1'(r) &= \frac{+(1+r^2)p_1(r)-3(1+r^2)q_1(r)}{r(1-r^2)} \, .
\label{6heq2}
\eeq
One can easily solve these coupled differential equations, and most generally the solutions are
\beq
p_1(r) & = c_1 (3-4r^2+r^4) + c_2 \frac{(1-r^4)(1-4r^2+r^4)}{r^4} \, , 
\\
q_1(r) & = c_1 (1-r^2)^2 - c_2 \frac{(1-r^2)^2(1+r^4)}{r^4} \, . 
\eeq
Now due to regularity at $r=0$, we should set $c_2=0$. Then $c_1$ can be set to any non-zero value by rescaling $\ep$. We choose $c_1=1/8$, since in that case our $\ep$ is conveniently identified with $\alpha$ in \eqref{uv6a}-\eqref{uv6d}.
This fixes $s_2$ as well, through \eqref{fors2}.
\beq
s_2(r) = \left(3-r^2\right)
   \left(1-r^2\right)^2
   \left(1+r^2\right)/64 . 
\eeq

So far we have considered only scalar field excitations. At one higher order of each BPS equation in 
\eqref{bps6a}--\eqref{bps6d}, we have four coupled first-order differential equations for $a_2,p_3,q_3,\s_4$. By themselves it is rather hard to find general solutions. But here the algebraic constraint \eqref{algcon} comes in handy. At ${O}(\ep^2)$, it gives an algebraic relation between $a_2, p_3, q_3, s_4$. One can solve it for {\it e.g.} $a_2$, and substitute it back to differential equations. Then the coupled first-order differential equations for $p_3,q_3,s_4$ can be solved explicitly, and the results are 
\beq
p_3 & = \frac{4
   r^{16}-32 r^{14}+100 r^{12}-112
   r^{10}-491 r^8+214 r^6+2340
   r^4-1488 r^2+372}{6144 r^4}
   \nn\\
 &  +\frac{c_1 \left(r^8-4
   r^2+1\right)}{r^4}
   -\frac{c_2}{3} 
   \left(r^4-3\right)-\frac{c_3}{2}
    \left(r^4-2r^2\right) \, , 
   \nn\\
q_3 & = \frac{2
   r^{16}-20 r^{14}+62 r^{12}+8
   r^{10}-713 r^8+335 r^6-46 r^4+744
   r^2-372}{6144 r^4} 
   \nn\\
&   +\frac{c_1
   \left(r^8-2 r^4+2
   r^2-1\right)}{r^4}
   -\frac{c_2}{3} 
   \left(r^4-1\right)+\frac{c_3}{2}
   \left(r^2-r^4\right) \, , 
   \nn\\
s_4 & = -\frac{-16
   r^{12}+192 r^{10}-1083 r^8+1484
   r^6+2895 r^4-4960 r^2+1488}{65536
   r^2} 
   \nn\\
&   -\frac{c_1 \left(r^8-3 r^4-4
   r^2+6\right)}{16 r^2}  +
\frac{ c_2}{48} \left(r^6-9
   r^2+8\right)+\frac{3 c_3}{64}
   \left(r^4-4 r^2+3\right)
   r^2 \, . 
\eeq
We again require that the scalars are finite at IR ($r=0$) and vanish at UV ($r=1$). These give us 
$$ c_1 = -31/512 , \quad c_3 = -(4096 c_2 + 1651)/3072 \, . 
$$
For the identification of $\alpha=\ep$ to hold up to this order, we demand 
$$
\lim_{r\ra 1} \frac{s_4}{(1-r)^2} = 0 \, , 
$$
and we can fix the coefficient $c_2= -99/256$. Substituting these values, we obtain
\beq
p_3(r) &=
   \frac{\left(r^2-1\right)^3
   \left(r^6-5 r^4+7
   r^2+9\right)}{1536} \, , 
   \nn\\
q_3(r) &=
   \frac{\left(r^2-1\right)^2
   \left(r^8-8 r^6+14 r^4+40
   r^2-47\right)}{3072} \, , 
   \nn\\
s_4(r) &=
   \frac{\left(r^2-1\right)^2
   \left(r^6-10 r^4+25
   r^2-16\right)}{4096} \, , 
   \nn\\
a_2(r) &= \frac{(r^2-1)^2(r^2-2)}{384} \, . 
\eeq

\begin{figure}[htbp]
\begin{center}
   \includegraphics[width=.45\textwidth]{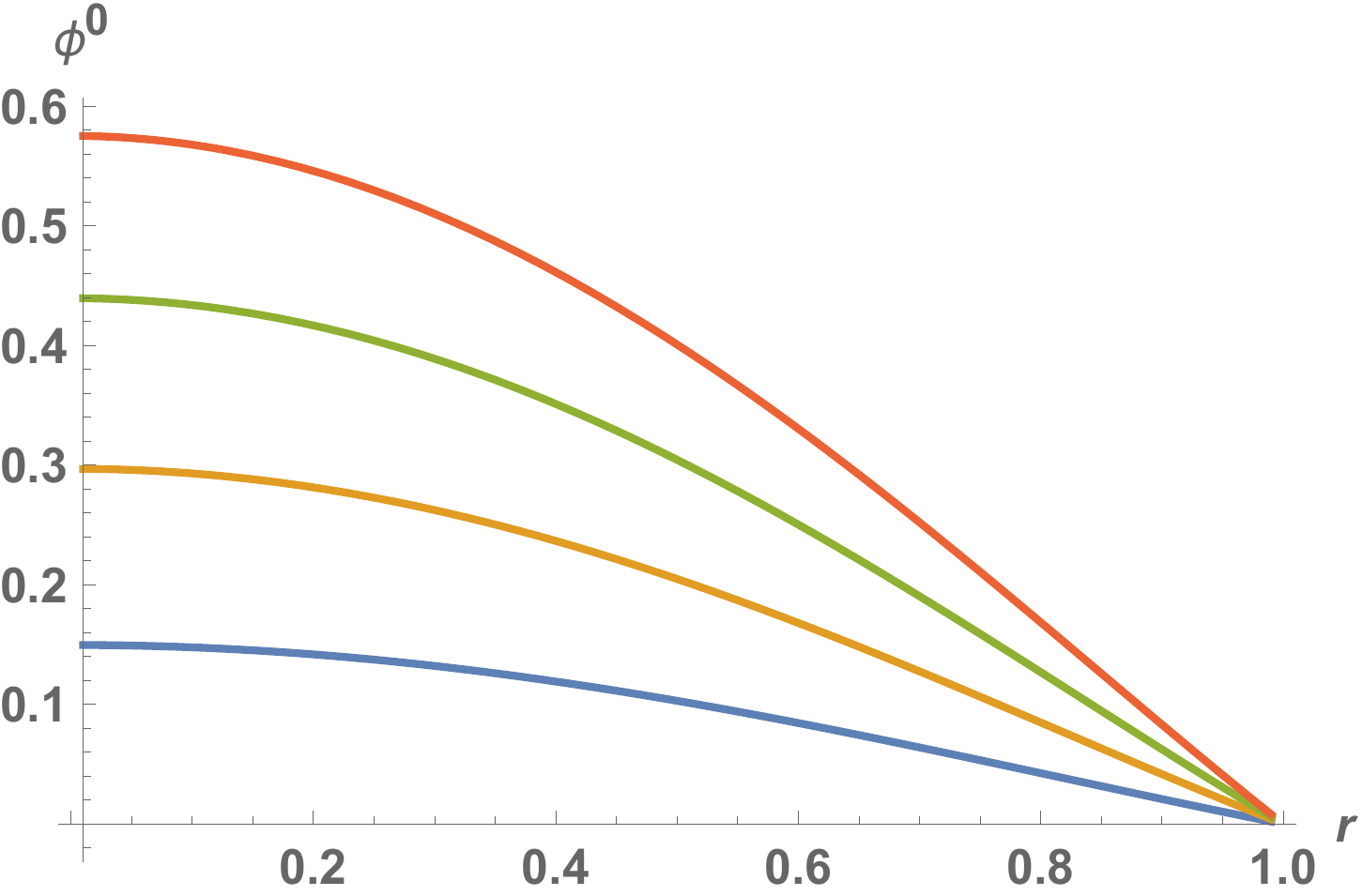}
   \quad
   \includegraphics[width=.45\textwidth]{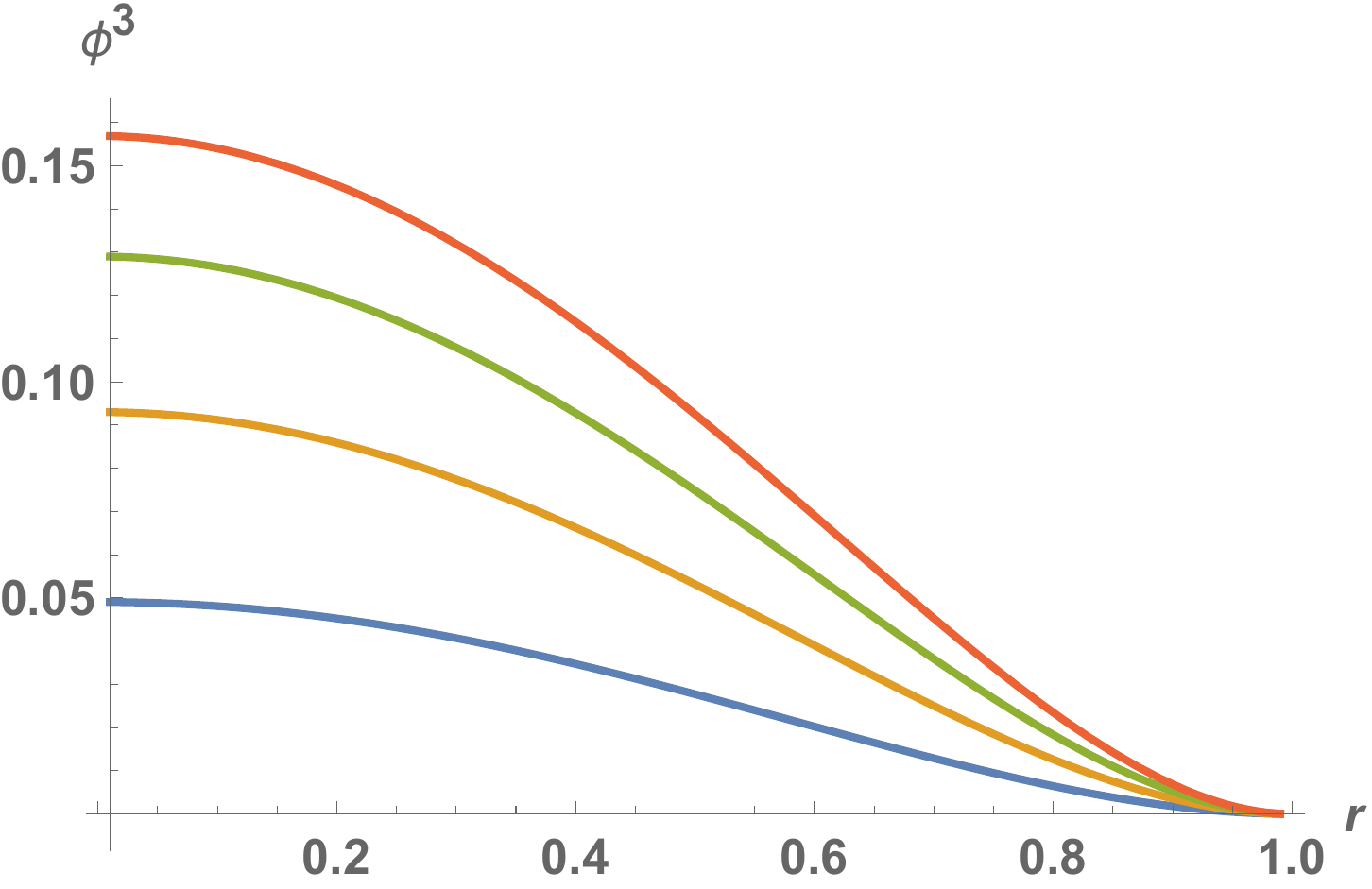}
   \\
   \includegraphics[width=.45\textwidth]{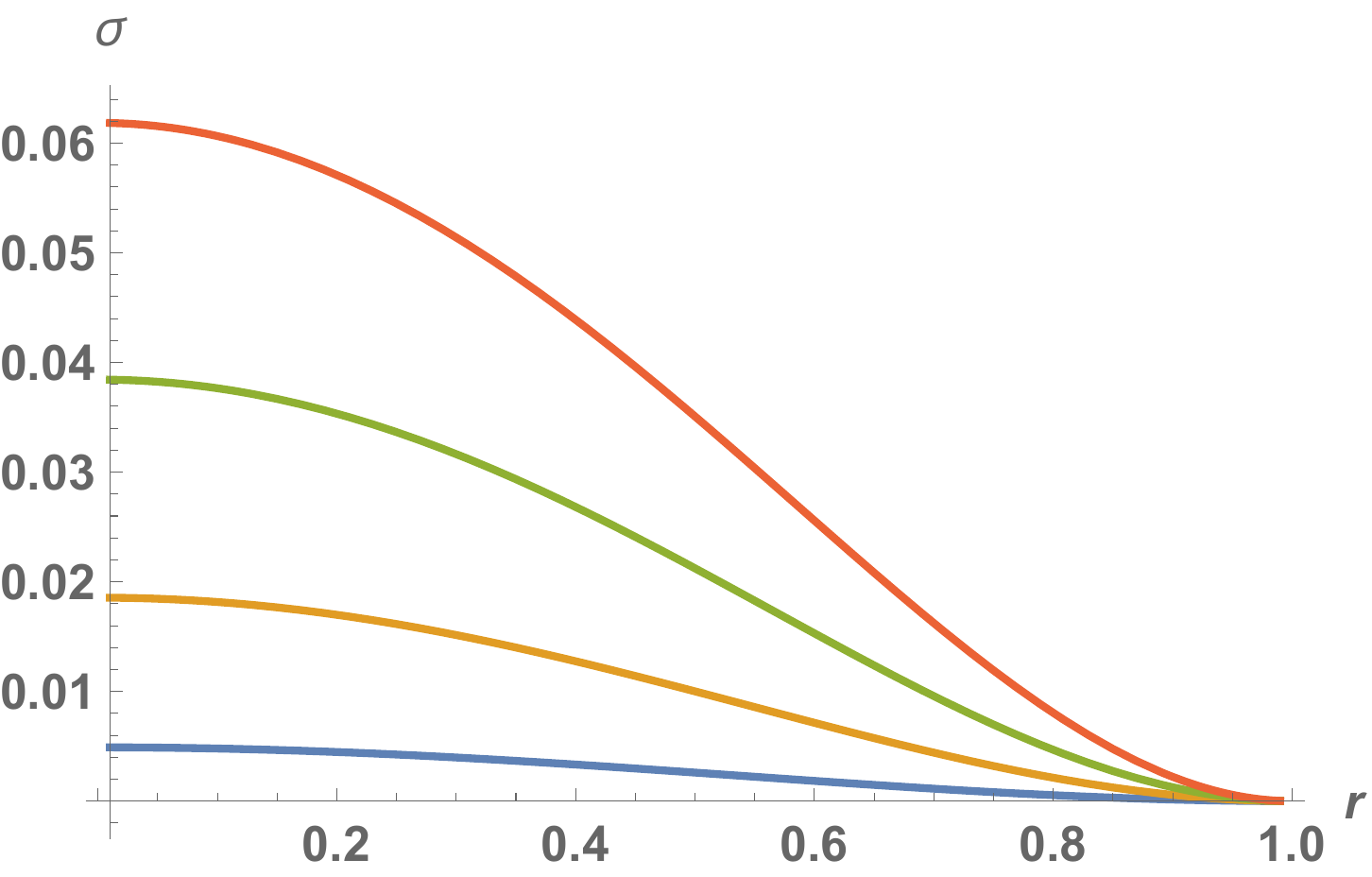}
   \quad
   \includegraphics[width=.48\textwidth]{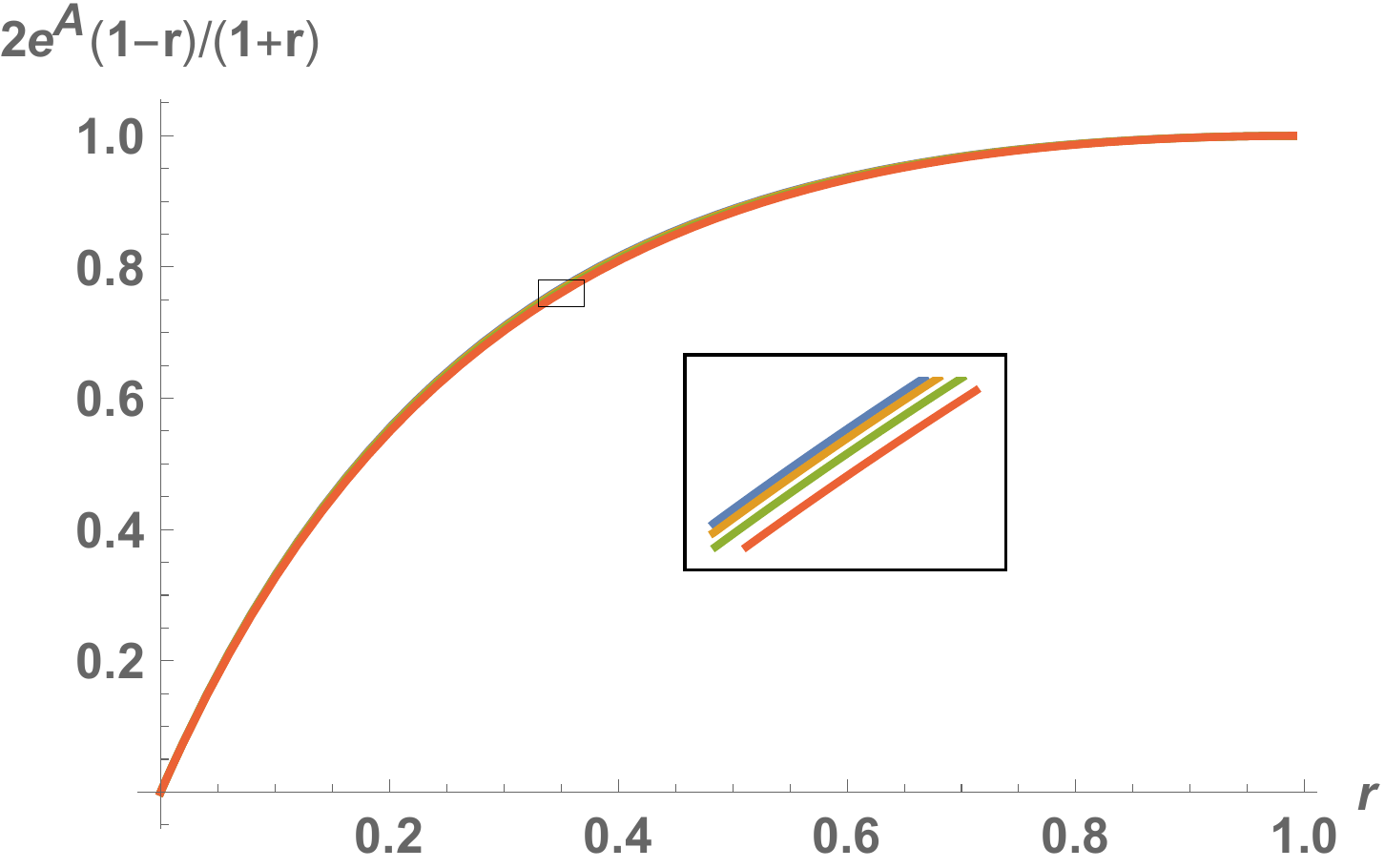}
   \end{center}
   \caption{Graphs of the BPS solutions based on our perturbative analysis. We choose $\ep=0.4,0.8,1.2,1.6$, and lower (higher) curves for $\phi^0,\phi^3,\sigma$ correspond to smaller (larger) values of $\ep$.}
    \label{fig:6sols}
\end{figure}

Obviously we can continue to do this to arbitrarily higher orders in $\ep$, in principle. At every order we obtain a system of BPS conditions for $a_{2n},p_{2n+1},q_{2n+1},s_{2n+2}$: one algebraic and four coupled first-order differential equations. With the help of the algebraic relation, we managed to solve the equations explicitly using {\it Mathematica}, up to $a_{24},p_{25},q_{25},s_{26}$. 
The solutions at each perturbative are always in polynomial form. This is easy to understand, because at higher orders we always have the same equations as \eqref{6heq1}--\eqref{6heq2}, now with inhomogeneous parts given in terms of lower-order solutions. Graphs of the solutions for a selection of $\ep=0.4,0.8,1.2,1.6$ are given in Fig.\ref{fig:6sols}.
Below we present just the next order results, so that the readers may feel how it will continue. 
\beq
 p_5(r) &= \frac{\left(r^2-1\right)^5}
{1474560}   \left(9 r^{10}-75 r^8+220 r^6-44
   r^4-587
   r^2+513\right) \, , 
   \nn\\
q_5(r) &= \frac{\left(r^2-1\right)^2}{11796480} \left(27
   r^{16}-336 r^{14}+1796
   r^{12}-4704 r^{10}+5218 r^8-5264 
   r^6 \right.\nn\\ & \quad\left.
   +28164 r^4-58048
   r^2+33147\right) \, ,
   \nn\\
s_6(r) &= -\frac{\left(r^2-1\right)^3}{7864320} \left(5
   r^{12}-77 r^{10}+522 r^8-2194
   r^6+5825 r^4-8241
   r^2+5120\right) \, ,
   \nn\\
a_4(r)&=-\frac{
   \left(r^2-1\right)^4}{1474560} \left(55
   r^4-348
   r^2+232\right) \, .
\eeq
Once we have explicit results, it is straightforward to work out the relation between $\alpha,\beta$, and $f_k$. We have already arranged $\alpha=\ep$, so determining $\beta,f_k$ as functions of $\ep$ gives $\beta(\alpha),f_k(\alpha)$ as well. From \eqref{uv6b} and \eqref{uv6d}, we see
\beq
   e^{f_k} \beta &= \frac{4}{\ep} \lim_{r\ra 1} \left[\s(r)- \frac{3\ep^2}{8}\frac{(1-r)^2}{(1+r)^2}\right] \frac{(1+r)^3}{(1-r)^3} ,
   \nn\\
   e^{-f_k} &= \frac{1}{\ep} \lim_{r\ra 1} \frac{\phi^3(r)(1+r)^2}{(1-r)^2} .
\eeq
Substituting our perturbative results up to $a_{24},p_{25},q_{25},s_{26}$, we  find $e^{-f_k} = 2$ precisely, and 
\beq
\beta (\alpha) & = -4 \alpha -\frac{\alpha ^3}{2}+\frac{\alpha
   ^5}{32}-\frac{\alpha ^7}{256}+\frac{5 \alpha
   ^9}{8192}-\frac{7 \alpha
   ^{11}}{65536}+\frac{21 \alpha
   ^{13}}{1048576}-\frac{33 \alpha
   ^{15}}{8388608}
    +\frac{429 \alpha
   ^{17}}{536870912}
      \nn\\
&   -\frac{715 \alpha
   ^{19}}{4294967296}+\frac{2431 \alpha
   ^{21}}{68719476736}-\frac{4199 \alpha
   ^{23}}{549755813888}+\frac{29393 \alpha
   ^{25}}{17592186044416}+ \cdots , 
\eeq
which in fact agrees with the series expansion of $-4\alpha \sqrt{1+ \alpha^2/4}$. Substituting these into \eqref{fprime}, 
\beq
F(\alpha)-F(0) = \frac{\pi^2}{3G_6} \left[ 1 - \left(1+\frac{\alpha^2}{4} \right)^{3/2} \right] \, . 
\eeq

Now let us compare this to the field theory side computation. Using the localization formula of \cite{Kallen:2012va}, the authors of \cite{Gutperle:2018axv} computed the large-$N$ limit of the mass-deformed partition function. As a function of $\mu=N^{-1/2}m$ ($m$ is the mass of a hypermultiplet in the fundamental representation of $USp(2N)$ gauge field theory), 
\beq
F (\mu) = \frac{\pi}{135} \left( (N_f-1) |\mu|^{5} - \sqrt{\frac{2}{8-N_f}}(9+2\mu^2)^{5/2} \right) N^{5/2} . 
\eeq
Since in fact $\mu$ is an $O(N^{-1/2})$ quantity, the above result can be trusted only for the leading order correction, {\it i.e.} $O(\mu^2)$ term. Matching it against the gravity computation only allows us to establish a relation between $\alpha$ (from gravity side) and $\mu$ (from field theory side). We find $\mu/\alpha = \frac{3\sqrt{30}}{20} = 0.821584$. We note that the numerical analysis of \cite{Gutperle:2018axv} gave $0.81$ instead. \footnote{After the first version of this article was submitted, through private communications, we were informed by J. Kaidi that using improved numerical methods the authors of \cite{Gutperle:2018axv} had found the value was closer to 0.82. }

\section{$AdS_5$: $\cN=2^*$ theory in $D=4$}
\label{ads5}
Let us now turn to the archetype of AdS/CFT, {\it i.e.} $\cN=4$, $D=4$ super Yang-Mills theory and its mass deformations. We consider here adding mass terms to two of the chiral multiplets and keeping $\cN=2$ supersymmetry. This is usually called $\cN=2^*$ theories. On the field theory side, when there is $\cN=2$ supersymmetry one can apply the work of Pestun \cite{Pestun:2007rz} for localization computations. In the large-$N$ limit, the result for free energy gives \cite{Buchel:2013id}
\beq
F_{S^4} = -\frac{N^2}{2} (1 + m^2 a^2) \log \frac{\lambda (1+m^2 a^2) e^{2\gamma+\frac{1}{2}}}{16\pi^2} \, . 
\eeq
Here $m$ is the mass of the hypermultiplet in $\cN=2^*$ theory, and the theory is put on the $S^4$ with radius $a$. Because the Euler-Mascheroni constant $\gamma$ implies that this result as it stands is scheme-dependent,  the authors of \cite{Bobev:2013cja} suggested we consider
\beq
\frac{d^3 F_{S^4}}{d(ma)^3} = -2N^2 \frac{ma(m^2a^2+3)}{(m^2 a^2+1)^2} , 
\label{ads5loc}
\eeq
where $\gamma$ disappears. The next task is then to construct BPS solutions on supergravity side, turning on the scalar fields which are dual to the mass terms of the field theory. One evaluates the holographically renormalized action, and identify the parameter of the gravity solution which is dual to $m$, and check if $F(ma)$ from the gravity side computation also satisfies \eqref{ads5loc}. This assignment was undertaken in \cite{Bobev:2013cja} and the authors claimed agreement based on numerical solutions in supergravity.

Let us now look at the gravity side computation more closely.
We are given the maximal gauged supergravity in $D=5$, and consider a truncated action (in the Euclidean signature) with scalars corresponding to mass terms of the chiral multiplets. Here we re-visit the analysis in \cite{Bobev:2013cja}, and illustrate how our perturbative approach improves the gravity side computation. The truncated action contains three scalar fields 
$\eta,z, \tilde z$, which are to be associated with various mass terms on the gauge theory side. 
\beq
\begin{aligned}
L &= \frac{1}{16\pi G_5} \left[ -R + \frac{\partial_\mu \eta \partial^\mu \eta}{\eta^2}
+ \frac{4 \partial_\mu z \partial^\mu \tilde z}{(1-z\tilde z)^2} + V \right] \, , \label{ads5eq} \\
V & \equiv -\frac{4}{L^2} \left( \frac{1}{\eta^4} + 2\eta^2 \frac{1+z\tilde z}{1-z\tilde z}
+ \frac{\eta^8}{4} \frac{(z-\tilde z)^2}{(1-z\tilde z)^2}
\right) \, . 
\end{aligned}
\eeq

The field equations allow BPS property, and the first-order equations without the warp factor are
\beq
\label{zzteq}
\begin{aligned}
z' & = \frac{3\eta'(z \tilde z - 1)\left[ 2(z+\tilde z ) + \eta^6(z-\tilde z)\right]}{2\eta 
\left[ \eta^6 (\tilde z^2 - 1 ) +\tilde z^2 + 1 \right]} 
\\
\tilde z' & = \frac{3\eta'(z \tilde z - 1)\left[ 2(z+\tilde z ) - \eta^6(z-\tilde z)\right]}{2\eta 
\left[ \eta^6 ( z^2 - 1 ) + z^2 + 1 \right]} 
\\
(\eta')^2 &= \frac{\left[\eta^6(z^2-1)+z^2+1\right]\left[\eta^6(\tilde z^2-1)+\tilde z^2+1\right]}{9\eta^2 (z \tilde z - 1 )^2}
\end{aligned}
\eeq
The warp factor $e^A$ in the metric convention of $ds^2=e^{2A}(dr^2/r^2+ds^2_{S^4})$ is determined by the scalar fields in the following way. 
\beq
\label{etaeq}
\begin{aligned}
\frac{\eta'}{\eta} &= \frac{\left[ 1+z^2+\eta^6(z^2-1)\right](A'\mp e^{-A})}{2(1+z^2)+\eta^6(1-z^2)}
\\
\frac{\eta'}{\eta} &= \frac{\left[ 1+\tilde z^2+\eta^6(\tilde z^2-1)\right](A'\pm e^{-A})}{2(1+\tilde z^2)+\eta^6(1-\tilde z^2)}
\end{aligned}
\eeq
There is also an algebraic constraint,
\beq
e^{2A} = \frac{(z \tilde z - 1 )^2\left[\eta^6(z^2-1)+z^2+1\right]\left[\eta^6(\tilde z^2-1)+\tilde z^2+1\right]}{\eta^8 (z^2 - \tilde z^2)^2} \, . 
\eeq
One can check that the above six equations are consistent with each other and also with the second order field equations which are derived from the action \eqref{ads5eq}. 

For our purpose, we also need to analyze the UV expansion of BPS solutions. Translating the result of \cite{Bobev:2013cja} to the metric convention $ds^2 = d\z^2/\z^2 + e^{2f(\z)} ds^2_{S^4}$, for small $\z$ we have 
\beq
\label{ads5exa}
\begin{aligned}
e^{2f} &= \frac{1}{4\z^2} + \frac{1}{6} ( \mu^2 - 3) + {\cal O} ( \z^2 \log^2 \z ) 
\\
\eta &= 1 + \z^2 \left[ -\frac{2\mu^2}{3} \log \z + \frac{\mu(\mu+v)}{3} \right] + {\cal O} (\z^4 \log^2 \z)
\\
 ( z + \zt )/2 & = \z^2 (-2\mu \log\z  + v) + \CO ( \z^4 \log^2 \z) 
\\
 ( z - \zt )/2 & = \mp \mu \z \mp \z^3 \left[
-\frac{4}{3}\mu (\mu^2-3) \log z  + \frac{1}{3} \left( 2v (\mu^2-3) + \mu(4\mu^2-3)\right) \right] 
\\
&  
+ \CO ( \z^5 \log^2 \z )
\end{aligned}
\eeq
Here $\mu,v$ are integration constants, and if we further impose regularity at IR (where $e^{2f}\rightarrow 0$) there should be a one-parameter family of regular solutions. Namely, we expect to find a functional relation between $v$ and $\mu$. This parameter of course should be related to the mass parameter in $\CN=2^*$ theory. Based on numerical solutions, the authors of \cite{Bobev:2013cja} conjectured that  for regular solutions 
\beq
\begin{aligned}
v(\mu) &= \sum_{k=1}^\infty v_{2k-1} \mu^{2k-1} 
\\
&= -2\mu +\mu^3 + \frac{\mu^5}{2}+ \frac{\mu^7}{3}+ \frac{\mu^9}{4} + \cdots
\\
&=  - 2\mu - \mu \log ( 1- \mu^2) \, . 
\end{aligned}
\label{ads5v}
\eeq
Through holographic renormalization, the free energy of the $\CN=2^*$ theory on $S^4$ should be given in terms of $v(\mu)$, and according to \cite{Bobev:2013cja}
\beq
\frac{d^3F}{d\mu^3} = - N^2 v''(\mu) \, . 
\eeq
When one substitutes \eqref{ads5v} into this, one gets $F'''(\mu) = - 2N^2 \mu(3-\mu^2)/(1-\mu^2)^2$ and it is the same as the field theory computation \eqref{ads5loc} if we identify $\mu = \pm i ma$. 

Let us now treat the BPS equations perturbatively. We use a similar expansion in $\ep$, which will be identified with $\mu$. 
\beq
\begin{aligned}
z(r)&= \sum_{k=1}^\infty \ep^k z_k (r) 
, \quad 
\tilde z (r) = \sum_{k=1}^\infty \ep^k {\tilde z}_k (r) , 
\\ 
\eta (r) &= \sum_{k=2}^\infty \ep^k \eta_k (r), 
\\
e^{A(r)} &= \frac{2r}{1-r^2} \left( 1 + \sum_{k=2}^\infty \ep^k a_k (r) \right).
\end{aligned}
\eeq
The UV expansion in \eqref{ads5exa} in fact implies that we may set $z_{2k}={\tilde z}_{2k}=\eta_{2k-1}=0$ for $k=1,2,3,\cdots$. From the leading nontrivial order of \eqref{etaeq}, one derives 
\beq
\eta_2  = \frac{{\tilde z}_1^2-r^2 z^2_1 }{3(1-r^2)} , 
\quad
\eta'_2 = \frac{4r(z^2_1- {\tilde z}_1^2)}{3(1-r^2)^2} \, . 
\label{eta2}
\eeq
Substituting these into the (leading nontrivial order of) first two equations of 
\eqref{zzteq}, we obtain
\beq
r(1-r^2) z'_1 +3 z_1 +  {\tilde z}'_1 &= 0 , 
\\
(1-r^2) {\tilde z}'_1 + r (z_1 + 3 {\tilde z}'_1 ) &=0 . 
\eeq
The general solution of the above equation is 
\beq
z_1 &= \frac{c_1(1-r^2)^2}{r^3} + \frac{c_2(1-r^2)}{r^3}
\left[ 2r -(1-r^2) \log \left(\frac{1+r}{1-r} \right)
\right] \, , 
\\
{\tilde z}_1 &= \frac{c_1(1-r^2)^2}{r} - \frac{c_2(1-r^2)}{r}
\left[ 2r +(1-r^2) \log \left(\frac{1+r}{1-r} \right)
\right] \, . 
\eeq
Again due to regularity at $r=0$, we set $c_1=0$. And in order to identify $\ep$ with $\mu$ in \eqref{ads5v}, we need to choose $c_2 = -1/8$. Then $\eta_2$ can be determined now using \eqref{eta2}. Note that this leading-nontrivial-order result is consistent with the conjecture \eqref{ads5v}, {\it i.e.} $v'(0)=v_1=-2$.

\begin{figure}[htbp]
\begin{center}
   \includegraphics[width=.48\textwidth]{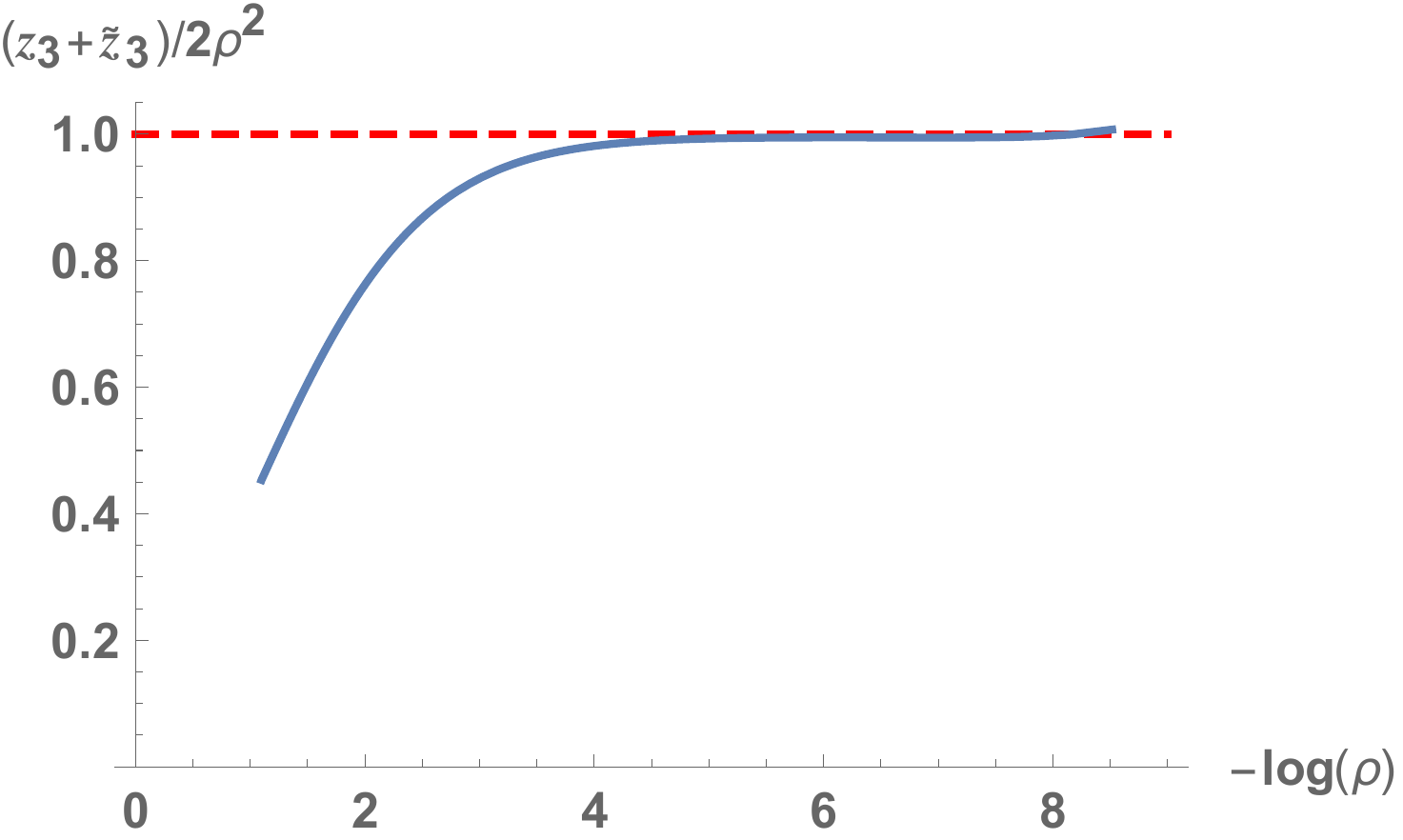}
   \quad
   \includegraphics[width=.48\textwidth]{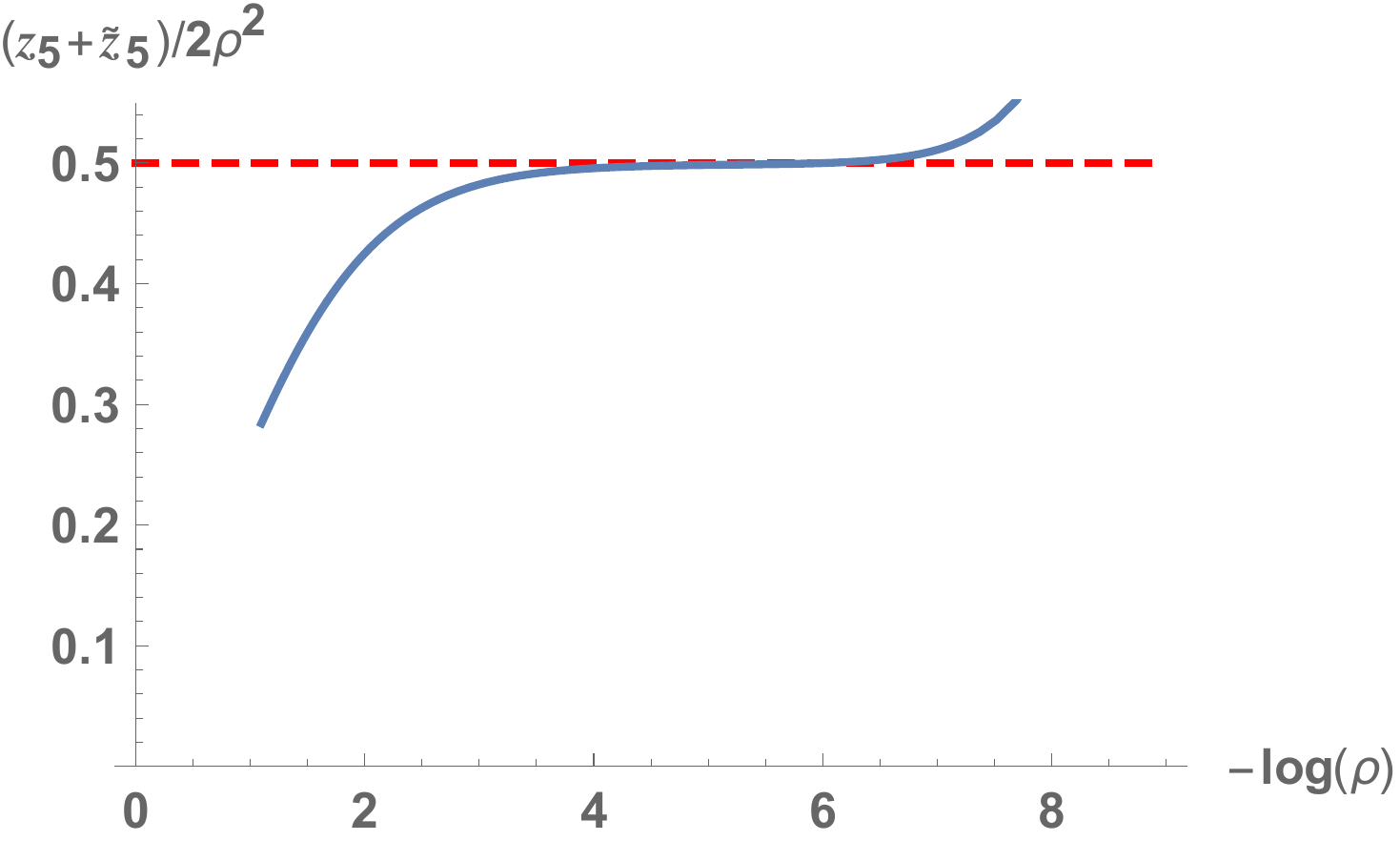}
   \\
   \includegraphics[width=.48\textwidth]{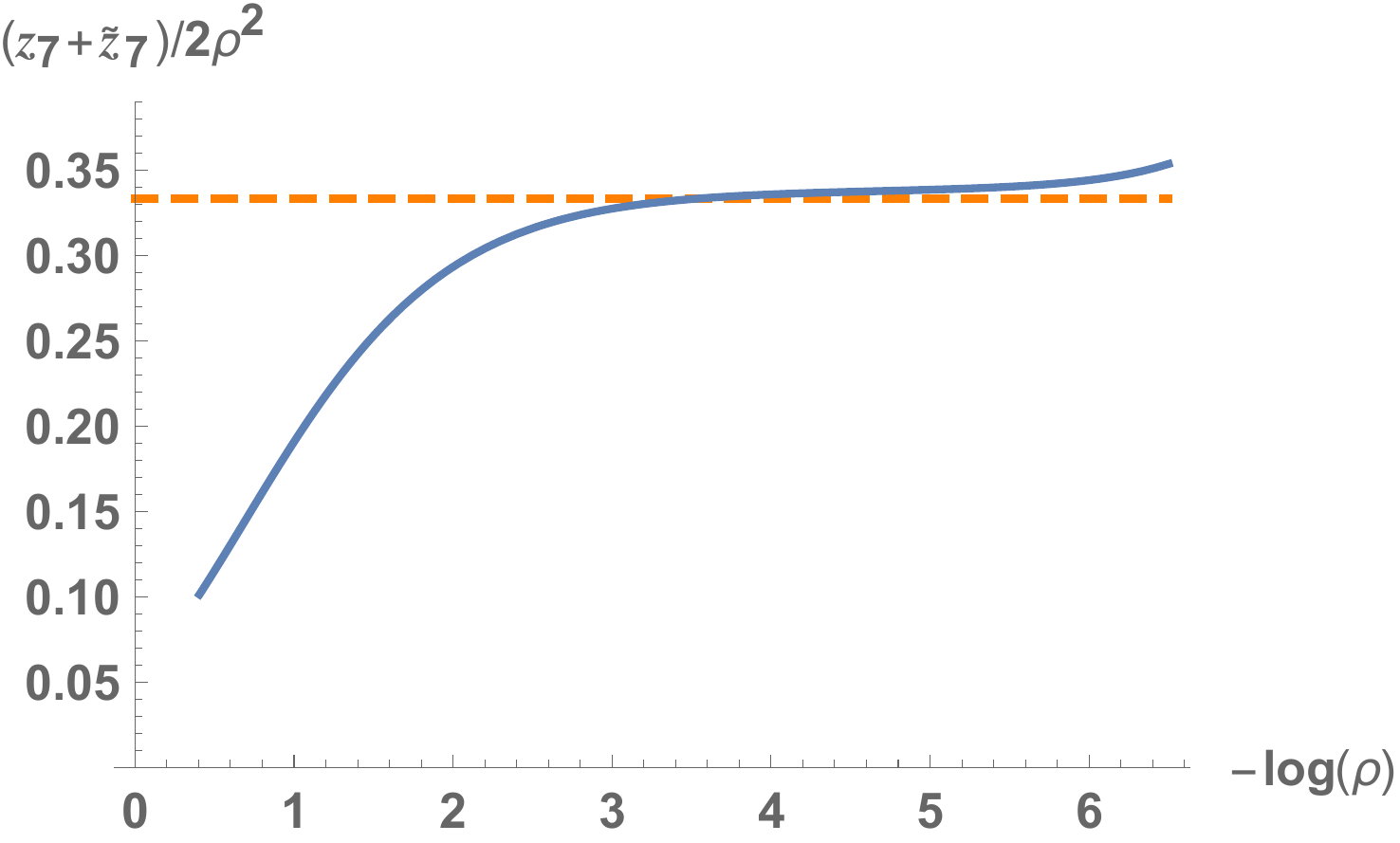}
   \quad
   \includegraphics[width=.48\textwidth]{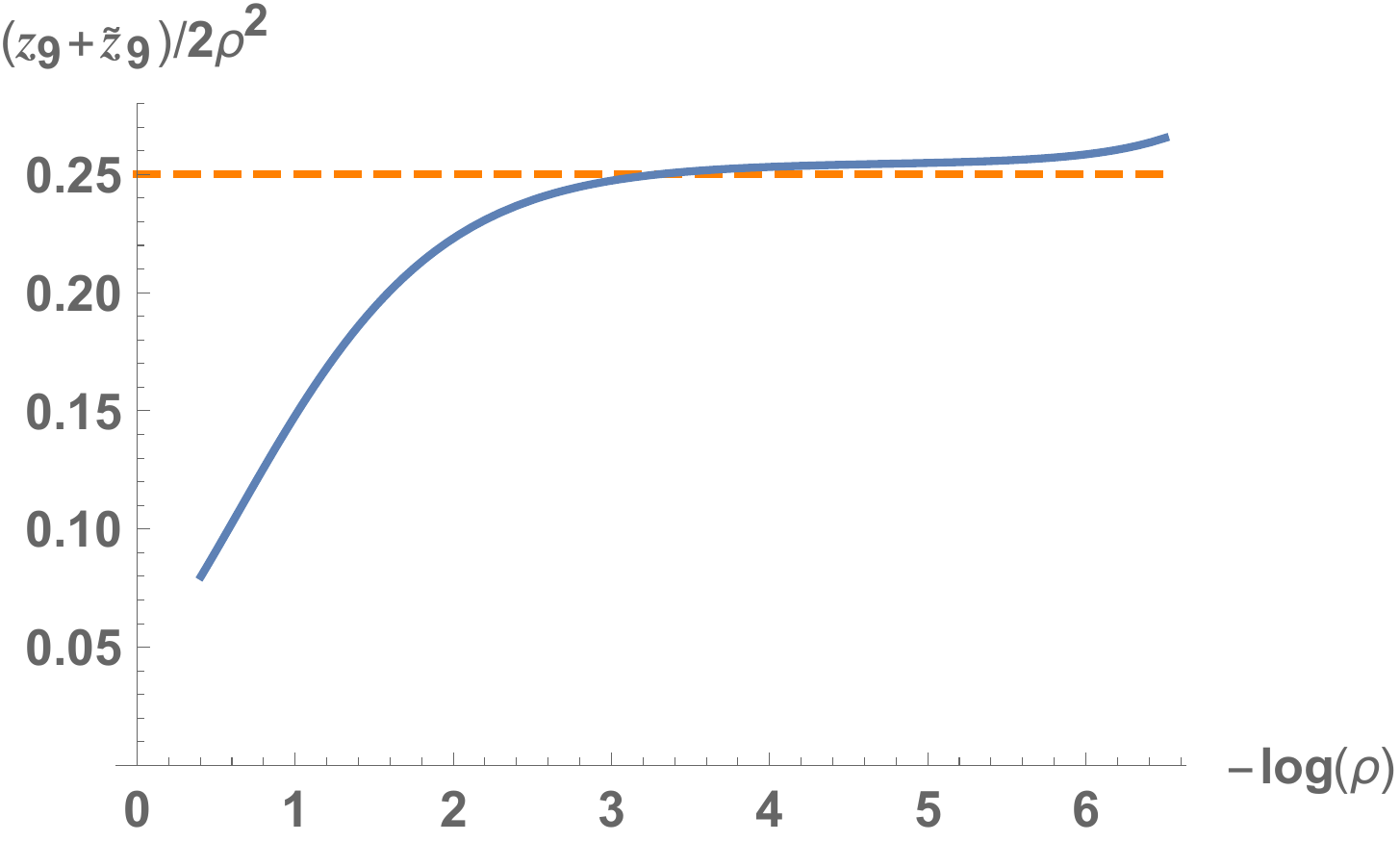}
   \\
   \includegraphics[width=.48\textwidth]{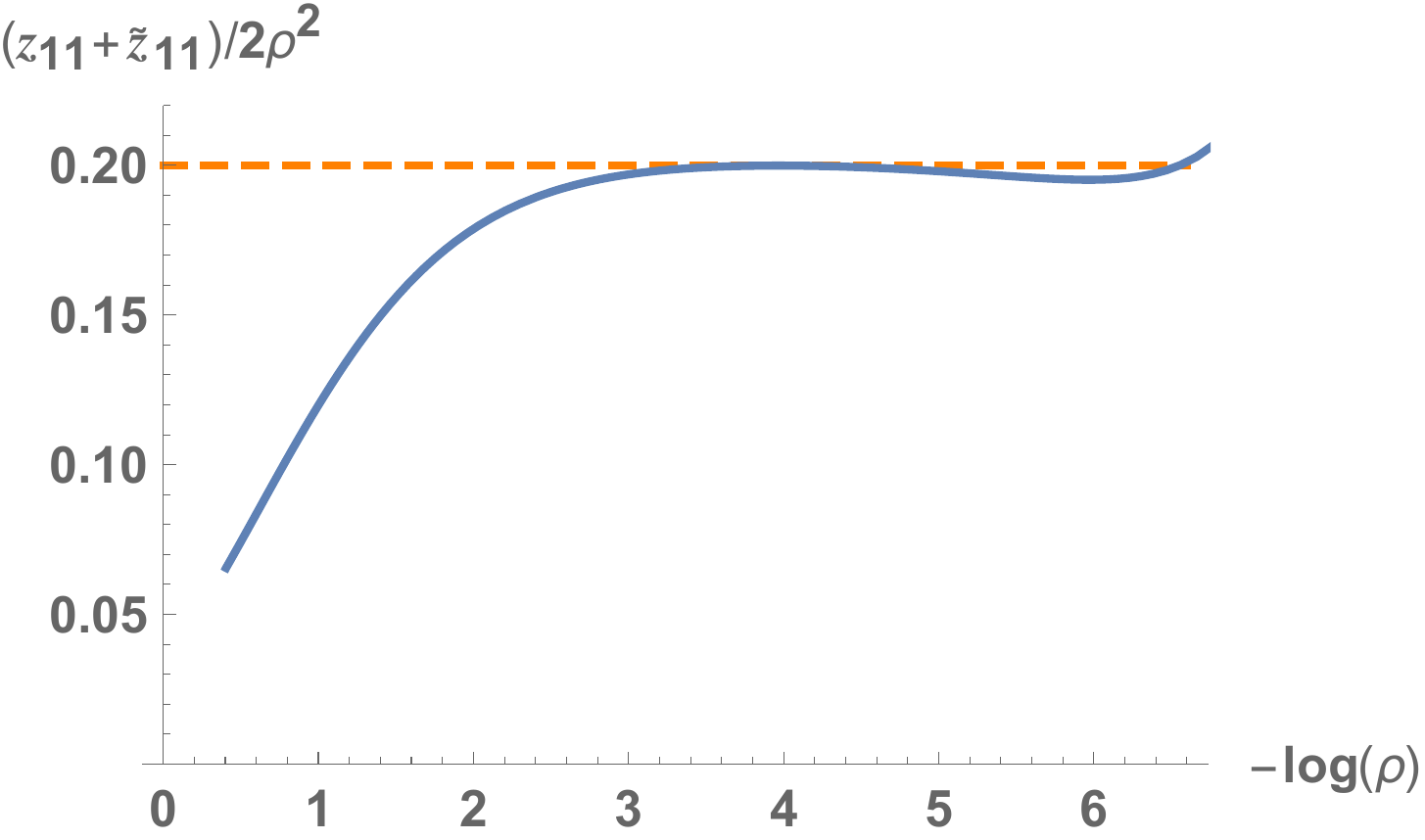}
   \quad
   \includegraphics[width=.45\textwidth]{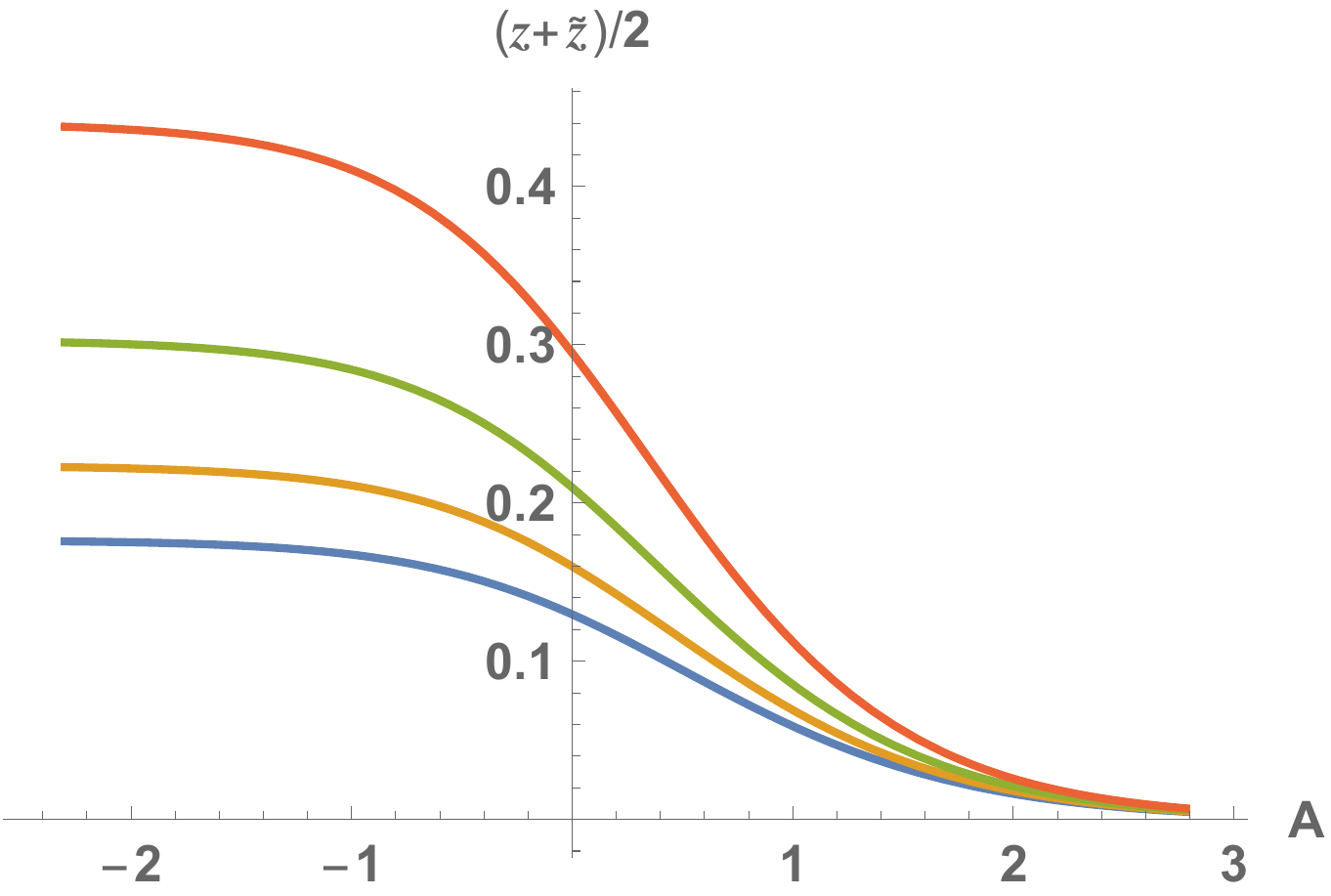}
   \end{center}
   \caption{Graphs of the BPS solutions based on our perturbative analysis. From the top-left one to the down-left one, we illustrate how our perturbative solution via series expansion agrees with the prediction $v_{2k+1}=1/k$ for $k\ge 1$. The horizontal axis in these plots represent $\log (1/\rho)$, and the vertical axis denotes $\frac{(z_{2k+1}+{\tilde z}_{2k+1})}{2\rho^2}$, where $\rho=(1-r)/(1+r)$. Because the series expansion is not analytic at $r=1$, the desired limits can be extracted at some value of $r$ close to 1. The bottom-right plot shows $(z+{\tilde z})/2$ as function of warp factor $A$, for several representative values of $\ep$. We choose $\ep=0.4,0.8,1.2,1.6$, and lower (higher) curves correspond to smaller (larger) values of $\ep$.}
    \label{fig:5sols}
\end{figure}

In principle, one may proceed to higher orders in $\ep$ just as we did in previous sections. However, the terms containing $\log$ in $z_1,{\tilde z}_1$  contribute to the in-homogeneous part of the differential equations for $a_2, z_3,{\tilde z}_3,\eta_4$, and their explicit integrations become rather messy. Instead of pushing analytic computation to higher orders, we have solved the system of coupled linear differential equations using a series expansion in $r$. In order to determine the integration constant, we demand regularity at $r=0$ (IR) and $r=1$ (UV). Without explicit integration results, a series expansion provides only approximate values for the integration constants, and the error affects the coefficients of all orders. We repeated the same calculation with different truncation lengths. Of course we have better agreement if we keep more terms. For concreteness we report here the case of using 500th order polynomials for $a_2, z_3,{\tilde z}_3,\eta_4$. We note that one can read off $v'''(0)$ by evaluating
\beq
v_{3} = \lim_{r\ra 1} \frac{(z_{3}+{\tilde z}_{3})(1+r)^2}{2(1-r)^2}  . 
\label{v3tolimit}
\eeq
Since the solutions are not analytic at $r=1$, we expand at $r=0$ and the radius of convergence for the series is 1. Because of numerical error, the numerator of the right-hand-side expression in \eqref{v3tolimit} is not exactly zero at $r=1$. The limiting value of \eqref{v3tolimit} can thus be extracted when we substitute a value which is close, but not too close, to 1 for $r$.  
A plot of this expression as a function of $\log\tfrac{1+r}{1-r}$ is given in Fig.\ref{fig:5sols}. We see that the plot approaches the desired value, 1, roughly for $4\lesssim \log\tfrac{1+r}{1-r}\lesssim 8$, {\it i.e.} $0.965\lesssim r \lesssim 0.999$.
For the next order of perturbation, we utilize the series expansion results for $a_2,z_3,\tilde{z}_3,\eta_4$ as an input to the in-homogeneous part of equations for $a_4,z_5,\tilde{z}_5,\eta_6$. From the plot of $\frac{(z_{5}+{\tilde z}_{5})(1+r)^2}{2(1-r)^2}$ we again see its value is close to the conjectured value of $1/2$, for $r$ close to $1$. We provide plots from our approximate solutions for $z_k+{\tilde z}_k$, for $k=3,5,7,9,11$ in Fig.\ref{fig:5sols}. Although they are not analytic results, our perturbative approach obviously lends support to \eqref{ads5v}.
\section{Discussion}
\label{discussion}
In this paper we have re-visited three problems in non-conformal precision holography. Most of the papers on this topic use numerically constructed solutions to supergravity field equations. At first sight it is understandable, since the BPS equations with scalar fields in supergravity are generically quite involved and analytic solutions are rarely  available. 

We suggest to use a perturbative technique instead. The AdS vacuum (for our purpose they are in Euclidean signature, {\it i.e.} hyperbolic space) is treated as $O(\ep^0)$ solution. Then we consider small departure from there, and solve iteratively, order-by-order in $\ep$. We see that at $O(\ep)$ we obtain equations which are not fully linearized yet, but are homogeneous. They are significantly simpler than the original equations, and we can solve them explicitly for all the examples we have considered in this paper. Then we substitute this result into $O(\ep^2)$ equations which are first-order and linear. Using elementary technique we can thus reduce the problems at $O(\ep^2)$ or higher, to just ordinary integrals. For $AdS_4$ and $AdS_6$ we have seen that the computation can be repeated practically to any higher orders. However for $AdS_5$ the integrand contains $\log$ terms and explicit integration gets quickly rather cumbersome. Of course we are familiar with the appearance of log terms in the study of odd-dimensional $AdS$: it is due to Weyl anomaly. We guess that in general also other examples of mass-deformed holography in $AdS_4$ and $AdS_6$ must be easier to solve, while for $AdS_5$ we do not expect to obtain all-order results. 


One may say that the method sketched in this paper is quite elementary, and ask why this was not tried before (except for references {\it e.g.} \cite{Bhattacharyya:2010yg,Alday:2014rxa,Alday:2014bta}, where perturbative method was applied to problems in different contexts of AdS/CFT). Or, why does this method work here at all? The reason why our method works nicely is related to the fact that the mass parameter can be treated as dimensionless, because we put our field theory on the sphere, not in Minkowski or Euclidean space. Since the free energy is dimensionless, the mass parameter should always appear in terms of $ma$ where $a$ is the radius of the sphere we put our field on. On the gravity side this is reflected on the fact that the boundary space is spherical in our ansatz, {\it e.g.} \eqref{metric4}.

Obviously our method can be applied to numerous other examples of AdS/CFT where we can consider adding mass to matter multiplets and work out the associated BPS equations. If there is still enough supersymmetry and localization is applicable, we can compare the gravity result with field theory. The most interesting one would be certainly mass-deformed ${\cal N}=4$ super Yang-Mills in four-dimensions. One can consider ${\cal N}=1^*$ theory, whose supergravity BPS equations are presented and solved numerically in \cite{Bobev:2016nua}. Results of our method will be reported elsewhere.
\section*{Acklowledgements}
The author thanks Hyojoong Kim, Se-jin Kim, Yein Lee for discussions and related collaborations. We also thank J. Kaidi for explaining his work \cite{Gutperle:2018axv} to us. This work was done partly during the APCTP workshop ``Strings, Branes and Gauge theories'', 16--25 July 2018, and we appreciate the hospitality. This work was supported by NRF grant 2015R1D1A1A09059301 and a grant from Kyung Hee University in 2016 (KHU-20160698). 
\bibliographystyle{JHEP}
\bibliography{ref}
\end{document}